\documentclass[
 preprint,
 eqsecnum,aps,prd,nofootinbib]{revtex4}
\usepackage[dvips]{graphicx}
\usepackage{floatfig}

\usepackage{amsmath}
\usepackage{enumerate}
\usepackage{mathrsfs}
\usepackage{amsfonts}

\def\be{\begin{equation}}
\def\ee{\end{equation}}
\def\bea{\begin{eqnarray}}
\def\eea{\end{eqnarray}}
\newcommand{\I}{{\mathscr I}} \newcommand{\tg}{{\tilde g}}
\newcommand{\tn}{{\tilde n}}
\newcommand{\tnabla}{{\tilde \nabla}}

\begin{document}
\initfloatingfigs
\title {Black Holes, AdS, and CFTs}

\author{Donald Marolf\footnote{\tt marolf@physics.ucsb.edu}}

\affiliation{Physics Department, UCSB, Santa Barbara, CA 93106, USA}

\begin{abstract}
This brief conference proceeding attempts to explain the
implications of the anti-de Sitter/conformal field theory (AdS/CFT)
correspondence for black hole entropy in a language accessible to
relativists and other non-string theorists.  The main conclusion is
that the Bekenstein-Hawking entropy $S_{BH}$ is the density of
states associated with certain superselections sectors,  defined by
what may be called the algebra of boundary observables.
Interestingly, while there is a valid context in which this result
can be restated as ``$S_{BH}$ counts all states inside the black
hole,'' there may also be another in which it may be restated as
``$S_{BH}$ does not count all states inside the black hole, but only
those that are distinguishable from the outside.'' The arguments and
conclusions represent the author's translation of the community's
collective wisdom, combined with a few recent results. For the
proceedings of the WE-Heraeus-Seminar: {\sl Quantum Gravity:
Challenges and Perspectives, dedicated to the memory of John A.
Wheeler.}
\end{abstract}

\maketitle

\tableofcontents

\section{Introduction}

In a classic set of lectures \cite{Wheeler}, John  Wheeler
emphasized the importance of the initial-value problem for quantum
gravity.  In particular, given any smooth set of initial data which
i) solves the gravitational constraints and ii) has small
curvatures, one expects the complete theory of quantum gravity to
contain a state which, in a suitable semi-classical limit,
approximates the spacetime generated by evolution from the given
initial data.  One expects this to hold even if the solution
develops singularities in both the far future and the far past.  In
such regions, quantum effects will be very important in describing
the physics of this state.   However, there will be a large region
of spacetime, including complete initial value surfaces, where the
semi-classical approximation holds.  It would therefore be a great
surprise if quantum gravity somehow contradicts the classical result
that such regions of spacetime can exist.

This argument continues to serve as an interesting point of
discussion for black hole physics, in particular in the context of
the anti-de Sitter/Conformal Field theory correspondence (AdS/CFT)
\cite{LargeN}.
  At the WE-Heraeus-Seminar: {\sl Quantum Gravity: Challenges and
Perspectives},  my charge was to discuss and explain the
implications of AdS/CFT  for  black hole entropy and unitarity, a
theme I will explore below using examples from \cite{Wheeler}. The
focus will be on the old question of whether the Bekenstein-Hawking
entropy counts {\em all} states inside black holes or only, in some
sense, those states which are distinguishable from the outside.
Interestingly, we will see that AdS/CFT suggests that there are
reasonable contexts realizing each of these two possibilities, but
that both cases are compatible with what may be called a ``unitary''
description of black hole evaporation.

\begin{figure}
\begin{center}
\scalebox
{.75} 
{
\includegraphics{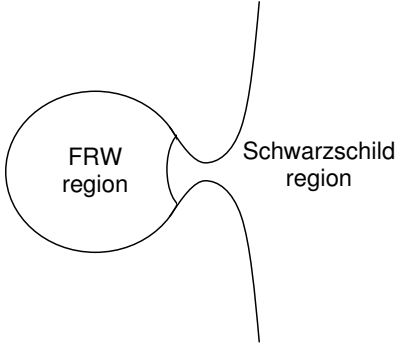}
} \caption{Moment of time-symmetry in a bag of gold spacetime.}
\label{gold}
\end{center}
\end{figure}

Let us begin by recalling what have come to be known as Wheeler's
``bag of gold" spacetimes\footnote{This terms seems to be in
frequent though informal use in the relativity community for the
spacetimes of figure \ref{gold}.  The actual connection to
\cite{Wheeler} is somewhat subtle, however, as Wheeler's original
use of the term was somewhat different.  Wheeler used the term ``bag
of gold'' in \cite{Wheeler} to refer to certain singularities that
arose in constructing time-symmetric initial data using conformal
methods when the conformal factor passed through zero.  The method
was intended for asymptotically flat spacetimes, but when such zeros
appeared they effectively pinched off part of the spacetime, tying
it up in a (singular) ``bag of gold.'' However, the same lectures
\cite{Wheeler} do introduce the spacetimes of figure \ref{gold},
though the emphasis is on the physics of the FRW side of the
Einstein-Rosen bridge as opposed to considering the FRW side as the
``inside'' of a black hole as we do here.}, in which a large $k=+1$
Friedmann Robinson Walker (FRW) universe is sewn onto the `back
side' of Kruskal extension of the Schwarzschild black hole spacetime
so that it is accessed by passing through the Einstein-Rosen bridge.
An embedding diagram for a time-symmetric slice is shown in figure
\ref{gold}. A detailed recent review of this construction can be
found in \cite{HR}.

Bag of gold spacetimes play an important role in discussions of the
meaning of black hole entropy.  Indeed, they typify one of the two
main classes of examples which suggest that black holes might have
an infinite number of internal states (even at finite Planck length
$\ell_p$) so that in particular the (finite) Bekenstein-Hawking
entropy $S_{BH}$ would not be simply the density of such states. The
point here is that, for a black hole of any fixed mass $M$,  the FRW
interior can be taken to be as large as one likes. Since the
time-symmetric slice of a $k=+1$ FRW universe of scale factor $a$
has energy density $\rho \sim 1/\ell_p^2a^2$, if this energy is in
the form of thermal radiation it contains an entropy $S \sim V
\rho^{3/4} \sim \left(\frac{a}{\ell_p}\right) ^{3/2}$.  Here the
symbol $\sim$ means that we have dropped constants of order one. We
see that the number of radiation states which can be placed inside
the FRW bag diverges as $a \rightarrow \infty$ so that, from the
perspective of the asymptotically flat region, an infinite number of
states can exist behind the black hole horizon\footnote{Other
examples that we consider to be in the same general class include
the Kruskal extension of the Schwarzschild black hole, in which the
2nd asymptotically flat region can be thought of as the limit of a
large FRW universe, and the ``monster'' initial data sets of
\cite{HR}. While the latter do not contain apparent horizons, they
contain a large amount of matter in an extremely deep throat-region
on the verge of gravitational collapse. It seems clear that collapse
must ensue on a timescale much too short for the matter to leave the
throat.  A black hole must then result and, while the initial data
surface contains no apparent horizon, part of this surface (and most
of the entropy) would nevertheless lie behind the event horizon and
in that sense be inside the black hole, as in the bag of gold case.
\label{foot}}.

It is interesting to compare the bag of gold spacetime with the other class of examples typically used to argue that black holes might contain an infinite number of internal states.  In this second example, one starts with a black hole of given mass $M$, considers some large number of ways to turn this into a much larger black hole (say of mass $M'$), and then lets that large black hole Hawking radiate back down to the original mass $M$.  Unless information about the method of formation is somehow erased from the black hole interior by the process of Hawking evaporation, the resulting black hole will have a number of possible internal states which clearly diverges as $M' \rightarrow \infty$.  One can also arrive at an arbitrarily large number of internal states simply by repeating this thought experiment many times, each time taking the black hole up to the same fixed mass $M' > M$ and letting it radiate back down to $M$.  We might therefore call this the `Hawking radiation cycle' exmaple.  Again we seem to find that the Bekenstein-Hawking entropy does not count the number of internal states.

I mentioned above that I will be interested in making contact with
AdS/CFT.  One of the interesting aspects of AdS/CFT is that the
entropy of the dual CFT is finite and agrees with the
Bekenstein-Hawking entropy $S_{BH}$.  This then provides contrasting
evidence that, in some sense, the Bekenstein-Hawking entropy
$S_{BH}$ {\em does} count the full set of black hole states in the
corresponding theory of quantum gravity.  We will review this
evidence below in section \ref{entropy}, but this foreshadowing
motivates a closer examination of the two examples above.  Is there
any way that they might be reconciled with such a finite-dimensional
space of black hole states?   Both examples admit ready
generalizations to the AdS context: adding a negative cosmological
constant to the bag of gold simply results in a $k=+1$ $\Lambda$-FRW
spacetime attached to the back side of an AdS-Schwarzschild black
hole, and adding a negative cosmological constant to the Hawking
radiation cycle example is straightforward when the black hole of
interest is very small compared to the AdS scale.  When the black
hole is larger than the AdS scale it becomes thermodynamically
stable and does not naturally radiate back down to a smaller black
hole on its own.  However, as we will discuss in section \ref{BO},
one can nevertheless remove the radiation by acting with certain
operators which are readily mapped to the CFT, draining away the
Hawking radiation ``by hand'' until the black hole evaporates back
to its original mass.

One notices two important differences between the bag of gold
example and that of the Hawking radiation cycle. The first is that
the bag of gold construction involves no Hawking radiation or other
quantum processes (except for counting the entropy of thermal
radiation).  Thus, while the Hawking radiation cycle example might
be reconciled with the view that $S_{BH}$ counts the number of
internal states if some mechanism could be found by which Hawking
radiation would erase the relevant information from the black hole
interior, such a reconciliation is not possible for bags of gold.

\begin{figure}
\begin{center}
\includegraphics{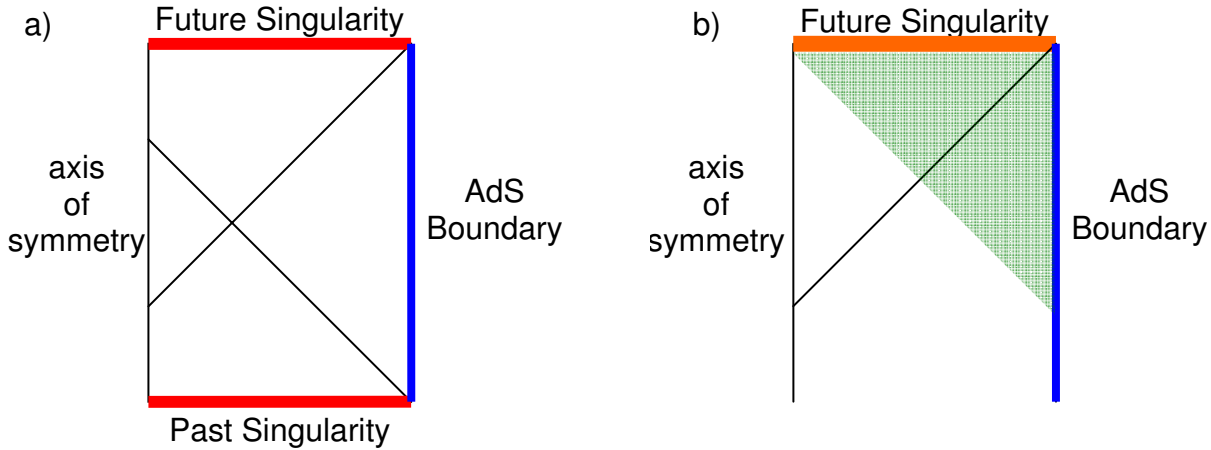}
 \caption{Rough conformal diagrams for a) an AdS bag of gold
spacetime and b) an asymptotically AdS black hole made by throwing
in matter (shaded) through the AdS boundary.} \label{conformal}
\end{center}
\end{figure}

The other difference is the dramatically different character of the
two examples at early times. The bag of gold spacetime contains a
past singularity, as do the related examples of footnote \ref{foot},
while the Hawking radiation cycle example can be formed from smooth
initial conditions in an asymptotically flat spacetime.  A similar
statement can be made about the AdS case. There one may begin the
Hawking radiation cycle example with completely empty anti-de Sitter
space and proceed by throwing in matter through the AdS boundary. In
contrast, there is no obvious way to construct a bag of gold
spacetime by simply manipulating perturbative excitations near the
AdS boundary\footnote{Though one could, of course, imagine tossing a
black hole with a bag of gold in through the boundary.}. See figure
\ref{conformal} for a comparison. As we will discuss below, one may
therefore interpret these two examples as belonging to different
superselection sectors of the theory associated with the algebra of
boundary observables.

Below, we first summarize the essentials of AdS/CFT in section
\ref{BO}.  For simplicity, we focus on the particular case where the
CFT lives on the Einstein static universe spacetime.  The main goal
is to explain the notion of boundary observables in the
asymptotically AdS quantum gravity theory, and we do so through the
example of the boundary stress tensor.  We also attempt to state the
sense in which the conformal field theory is ``dual'' to a quantum
gravity theory with asymptotically anti-de Sitter boundary
conditions.  We then briefly discuss issues of unitarity and
information loss in section \ref{unitarity} before turning to our
main discussion of entropy in section \ref{entropy}. The emphasis in
section \ref{entropy} is on making precise, conservative statements.
We then further discuss the implications of these statements in
section \ref{disc}.  By necessity, section \ref{disc} is less
precise, but we hope that the desire to flesh out the details will
lead to interesting future research.  In all sections, we attempt to
use language accessible to relativists and practitioners of quantum
field theory in curved spacetime.

\section{Boundary observables and AdS/CFT}

\label{BO}

As is by now well known, arguments from string theory \cite{LargeN}
suggest that at least certain theories of asymptotically anti-de
Sitter quantum gravity are in some sense ``dual'' to the large $N$
limit of certain local conformal field theories.  Here $N$ is
some measure of the number of CFT fields involved, which in typical
cases can be thought of as the rank of a gauge group associated with
some gauge field.   The correspondence furthermore identifies $N$
with some positive power of $\ell_{AdS}/\ell_p$; i.e.,
$\ell_{AdS}/\ell_p \propto N^\alpha$, where $\ell_{AdS}$ is the AdS
scale and the particular value of $\alpha > 0$ depends on the
particular duality being considered.  Many different such dualities
are believed to hold, associated in string theory with D3-branes,
M2-branes, M5-branes, or with bound states including several types
of branes. Each duality relates a CFT to some string or M-theory with
boundary conditions which require spacetimes to be asymptotically
AdS${}_d \times X$, where $X$ is some compact manifold and the
details of $d$, $X$, and the strng theory vary with the duality.

However, we will avoid discussing the details of any particular case
here.  Such details are simply not relevant to our discussion below.
Instead, we focus on features common to all AdS/CFT correspondences.
In particular, we will ignore the manifold $X$.  Effectively, one
may suppose that a Kaluza-Klein reduction has been performed to
replace fields on AdS${}_d \times X$ with an infinite tower of
massive fields living just on AdS${}_d$.   For definiteness, we
focus on the case where the CFT is defined on the Einstein static
universe spacetime $S^{d-2} \times \mathbb{R}$ below.

Now, we could begin by stating that the CFT Hilbert space is
isomorphic to a Hilbert space on the quantum gravity side of the
correspondence. However, this comment contains little information
since any two separable infinite-dimensional Hilbert spaces are
already well-known to be isomorphic.  Another statement  that might
be made is that the algebra of CFT operators is isomorphic to an
algebra of quantum gravity operators, and that the above isomorphism
of Hilbert spaces can be thought of as an isomorphism of
representations of this operator algebra.  However, this again
provides little information since each of the trivial isomorphisms
between two separable Hilbert spaces readily defines an isomorphism
of the associated von Neumann algebras, and also of the desired
representations.   We must clearly be more explicit to give a
non-trivial statement of the correspondence.

What gives AdS/CFT content is that we understand something about how
physically interesting operators map between the AdS gravity theory
and the conformal field theory.   Rather than attempt to state this
in great generality, it suffices for our purposes to explain a
particularly important example.  The claim is that the local
stress-energy tensor $T_{ij}^{CFT}(y)$ of the CFT maps directly to
an object  $T_{ij}^{AdS}(y)$  which we will explain below.  This
$T_{ij}^{AdS}(y)$  is known as the  ``boundary stress tensor" of the
AdS quantum gravity theory \cite{HS,kraus}.  In contrast to the
above general statements, this claim contains {\em vast} amounts of
information. For example, from the stress tensor one can construct
the Hamiltonian, which on the CFT side has a non-degenerate ground
state.  Thus AdS/CFT states that the quantum gravity theory has a
corresponding ground state and that, in that state, arbitrary
correlation functions (i.e., $n$-point functions, for all $n$) of
$T_{ij}^{AdS}(y)$ agree with the corresponding CFT vacuum
correlators of $T_{ij}^{CFT}(y)$.  This is now a very specific
prediction.

Let us take a moment to remind the reader what is meant by the AdS
boundary stress tensor $T_{ij}^{AdS}(y)$.  This also provides an
opportunity to state more precisely what we mean by asymptotically
AdS quantum gravity.  For our purposes here, it suffices to use an
extremely simple set of boundary conditions.  We consider spacetimes
$M$ such that:
\begin{enumerate}
\item One can attach a boundary $\I \cong R \times S^{d-2}$
to $M$ such that $\tilde M = M \cup \I$ is a manifold
with boundary.
\item
On $\tilde M$, there is a $(d-1)$-times continuously differentiable
metric $\tg_{ab}$ and a smooth function $\Omega$ such that $g_{ab} =
\Omega^{-2} \tg_{ab}$, with $\Omega = 0$ and \be \tn_a \equiv
\tnabla_a \Omega \neq 0 \ee at points of $\I$. We also require that
the metric $g_{(0)ij}$ on $\I$ induced by $\tg_{ij}$ is the Einstein
static universe, \be \label{g0} g_{(0)ij} \, dy^i dy^j = -dt^2 +
\ell_{AdS}^2 d \sigma^2, \ee where $d\sigma^2$ is the line element
of the unit sphere $S^{d-2}$ and $y^i$ are a set of coordinates on
the boundary.
\end{enumerate}

Given such an asymptotically AdS spacetime $(g, M)$, the
corresponding $\Omega$, and an extension of the $y^i$ into the bulk,
there is some diffeomorphism one can apply to $g_{ab}$  such that
the unphysical metric takes the ``Gaussian normal form''
 \be
\label{gFG} \tilde g_{ab} dx^a dx^b = \ell^2_{AdS} d \Omega d\Omega
+ \sum_{n \ge 0}   \Omega^n g_{(n)ij}   dy^idy^j  + O(\Omega^{d
+1}).
 \ee
The so called Fefferman-Graham coefficients $g_{(n)ij}$ for $n <
d-1$ are determined by $g_{(0)ij}$ and the Einstein equations,
though $g_{(d-1)ij}$ contains new information \cite{FG} which
defines the boundary stress tensor \cite{HS,kraus}:
 \be
  T_{ij}^{AdS}(y)
= \frac{d-1}{16 \pi G} g_{(d-1) ij}(y).
 \ee
We refer to \cite{HS,kraus} for a detailed explanation of why this
object is called the boundary stress tensor, though we comment
briefly that if $\xi^i_{boundary}$ is a conformal Killing field of
the Einstein static universe and $C$ is any Cauchy surface of $\I$,
the expression
 \be \int_C \sqrt{g_0} \ T_{ij}^{AdS} \xi^i_{boundary}
n_{C}^j,
 \ee
 with $n_C^j$ the unit future-directed timelike normal
to $C$ with respect to $g_{(0)ij}$, gives the conserved
charge\footnote{Though sometimes with a choice of zero-point
different from that typically chosen by relativists.  See e.g.
\cite{HIM,HIM2} for discussions of this point.} associated with the
asymptotic symmetry that acts on $\I$ via $\xi^i_{boundary}$.  In
particular, $T_{ij}^{AdS}(y)$ turns out to be essentially the
electric part of the Weyl tensor at $\I$ \cite{HIM}, which is known
to give the appropriate conserved charges \cite{AM,AD}.

Because the coefficients $g_{(d)ij}$ are defined in terms of the
Gaussian normal form (\ref{gFG}),  $T_{ij}^{AdS}(y)$ does not depend
on the extension of the coordinates $y^i$ into the bulk. Instead,
$T_{ij}^{AdS}(y)$ is a tensor on the boundary spacetime $\I$. As a
result, $T_{ij}^{AdS}(y)$ defines an {\em observable}.  By this we
mean that it is invariant under all gauge transformations of the
theory. We shall not go into the details of this story here, but
merely sketch an outline that should be familiar to most relativists
e.g., from the asymptotically flat context.  Beginning with the
usual gravitational symplectic structure (see e.g. \cite{WIWZ}), one
defines infinitesimal gauge transformations as tangent vectors to
the covariant phase space which yield degenerate directions of the
symplectic structure.  One then finds that diffeomorphisms are only
gauge transformations if they are generated by vector fields $\xi^a$
which are smooth on $\tilde M$, vanish on $\I$, and preserve the
above notion of asymptotically AdS spacetimes.   See e.g.
\cite{HIM,KSthermo} for details. Since gauge transformations act
trivially on $\I$, they also act trivially on any tensor on $\I$
such as $T_{ij}^{AdS}(y)$.  Thus we refer to $T_{ij}^{AdS}(y)$ as a
boundary observable. Note that, in contrast, diffeomorphisms of
$\tilde M$ which preserve the space of asymptotically AdS metrics
but which act non-trivially on $\I$ define asymptotic symmetries
which act non-trivially on observables.

In a similar way,  Fefferman-Graham-like expansions of any other
fields also define boundary observables.  The set of such
observables for all bulk fields is one of the primary ingredients of
the AdS/CFT dictionary.  The dictionary that maps such observables
to local CFT observables is understood in each example of AdS/CFT,
at least at leading order in the $1/N$ expansion.

Other CFT observables  (such as Wilson loops) can also be mapped to
AdS observables using more stringy ingredients (see e.g. \cite{WL}).
The details will not be important here, but we briefly mention that
they again lead to AdS observables associated with certain regions
of the boundary, and which transform in a natural way under
asymptotic symmetries. Furthermore, unless the regions $R_1$ and
$R_2$ on $\I$ associated with two such observables are connected by
causal curves, the corresponding CFT observables commute.  As more
observables are added to the dictionary, these important properties
are expected to hold in each case.  Thus the AdS/CFT dictionary
generally relates CFT observables to what may reasonably be called
boundary observables of the quantum gravity theory.

We will not need the details of any further such constructions
below, nor will it be necessary to give a precise statement of the
full set of boundary observables.    Instead, it will suffice to
suppose merely that there is some algebra of CFT observables ${\cal
D}_{CFT}$ and some algebra of quantum gravity observables ${\cal
D}_{AdS}$, for which this dictionary has been stated; i.e., we have
been given a particular bijective map $\phi_{AdS/CFT} : {\cal
D}_{CFT} \leftrightarrow {\cal D}_{AdS}$. The AdS/CFT conjecture is
the claim that, for every state $\rho_{CFT}$ in the CFT, there is
some state $\rho_{AdS}$ in the asymptotically AdS quantum gravity
theory such that the restriction\footnote{Here we use the language
of algebraic quantum field theory in which one thinks of a state as
a positive linear functional on an algebra. One may then restrict
any state to a subalgebra, e.g. to ${\cal D}_{AdS}$, the subalgebra
of AdS quantum gravity operators for which the details of the
dictionary have been stated.} of $\rho_{AdS}$ to ${\cal D}_{AdS}$
agrees with the image under $\phi_{AdS/CFT}$ of the restriction of
$\rho_{CFT}$ to ${\cal D}_{CFT}$.

As one gains more control over the AdS/CFT dictionary and the
algebra ${\cal D}_{AdS}$ becomes larger, the above claim becomes
increasingly powerful and the space of states $\rho_{AdS}$ which
might correspond to a given $\rho_{CFT}$ becomes smaller. In fact,
it is quite plausible that the sort of AdS boundary observables
defined above via Fefferman-Graham expansions are already dual to a
complete set of CFT operators.  As we will shortly explain, this
will be true if something like an ergodicity conjecture holds for
the CFT, which one might expect to hold due to its strongly
interacting nature.

To understand this last point, consider just the algebra of
observables generated by the CFT stress tensor $T_{ij}^{CFT}(y)$.
Since this algebra contains the Hamiltonian, it contains the
projection onto the CFT  vacuum.  But one expects to be able to
create rather general superpositions of energy eigenstates by using
a tensor test field $\sigma^{ij} : \I \rightarrow {\mathbb C}$ to
construct a unitary operator $U[\sigma^{ij}] = \exp\left( \int_\I
\sqrt{g_{(0)}} \sigma^{ij} T_{ij}^{CFT} \right)$ with which one may
act on the CFT vacuum.  Acting further with functions of the
Hamiltonian $H$ and taking linear superpositions plausibly generates
all states in the CFT Hilbert space\footnote{The only case in which
one expects this not to happen occurs when there is some symmetry
that leaves the stress tensor invariant; e.g., global charge
rotations.  In such a case one also needs to include in ${\cal
D}_{CFT}$ some operator not invariant under this symmetry.  This is
not difficult in specific examples and proceeds in direct analogy to
our discussion of  $T_{ij}^{CFT}(y)$ and $T_{ij}^{AdS}(y)$ above.}.
We shall therefore assume that ${\cal D}_{CFT}$ contains a complete
set of observables in our discussion below. (In any case, including
the Wilson loop observables mentioned above would certainly make
${\cal D}_{CFT}$ complete.)  It is then clear that two distinct CFT
states $\rho^1_{CFT}, \rho^2_{CFT}$ cannot both map to the same
state $\rho_{AdS}$.

On the other hand, we defer any discussion of whether each
$\rho_{CFT}$ has a unique $\rho_{AdS}$ to section \ref{disc}.
Nevertheless, it will be convenient to use the term ``AdS vacuum" to
refer to a particular state $\rho^0_{AdS}$ whose correlators agree
under $\phi_{AdS/CFT}$ with those of the CFT vacuum.  If there is
more than one such state, for now we simply pick one, say
$\rho^0_{AdS}$, and call it the AdS vacuum.  We may use this
$\rho^0_{AdS}$ to extend $\phi_{AdS/CFT}$ to a one-to-one map from
CFT states to AdS states: it maps the CFT vacuum to the AdS vacuum,
and it maps the action of ${\cal O}$ on the CFT vacuum to the action
of $\phi_{AdS/CFT}({\cal O})$ on the AdS vacuum.

As a brief aside, we note that while we have fixed $g_{(0)ij}$ to be
the metric on the Einstein static universe, one may in fact
generalize the discussion to any smooth Lorentz-signature metric on
which quantum field theory may be defined.  For any $g_{(0)ij}$, the
dual CFT is given by the same Lagrangian\footnote{Any
renormalization ambiguities associated with passing to a general
$g_{(0)ij}$ on the CFT side of the correspondence turn out to have
direct analogues on the AdS side.}
 as for the Einstein static universe,  but built instead from the metric $g_{(0)ij}$.  The Fefferman-Graham
expansion  (\ref{gFG})   and the construction of $T_{ij}^{AdS}(y)$
generalizes readily to this case, though when $d$ is odd an
additional logarithmic term must be added to (\ref{gFG}) for general
$g_{(0)ij}$.  The boundary stress tensor also receives contributions
from this logarithmic term.  Such a term was not included in
\eqref{gFG} because the coefficient of this logarithm happens to
vanish for the Einstein static universe; see e.g. \cite{HS,KSthermo}
for details.  For simplicity, we avoid considering such general
$g_{(0)ij}$ and confine ourselves to the Einstein static universe
case below.

Since they are dual to local fields in the CFT, one might expect
$T_{ij}^{AdS}(y)$-like boundary observables (defined by
Fefferman-Graham expansions of general bulk fields) to act much like
local fields themselves.  This does indeed turn out to be the case;
see e.g. \cite{Rehren}.  In particular, by smearing local boundary
observables with positive- and negative-frequency functions on the
boundary, one can define operators that act like creation and
annihilation operators \cite{Steve}.  It is precisely these
operators which may be interpreted as ``throwing particles into the
AdS space through the boundary,'' or as removing such particles
through the boundary.    This interpretation follows from a close
connection between such operators and deformations of the AdS
boundary conditions \cite{witten}.  For example, it turns out that
acting with $T^{AdS}_{ij}(y)$ is equivalent to deforming the
boundary metric $g_{(0)ij}$ at the event $y$.  The point here is
that a single state of the deformed theory can be used to construct
{\em two} states of the undeformed theory.  There is a ``retarded
state'' defined by noting that the two theories are identical to the
past of $y$, as well as a corresponding ``advanced state.''  The
advanced state is then the action of $T^{AdS}_{ij}(y)$ on the
retarded state, in what is essentially an example of the Schwinger
variational principle \cite{Schwinger,Bryce}, see \cite{Lorentz} for
details.

The fact that one may construct such boundary creation and
annihilation operators has two implications for our black hole
discussion.  First, it means that one may think of the Hawking
radiation cycle experiment from the introduction in terms of
applying boundary creation and annihilation operators in alternating
cycles and examining the resulting states. Second, it tells us much
about the vacuum state $\rho^0_{AdS}$.  Since the CFT vacuum is
annihilated by the CFT annihilation operators, the boundary
annihilation operators must also annihilate $\rho^0_{AdS}$.  Thus,
$\rho^0_{AdS}$ is a state for which no energy can be removed by
changing the boundary conditions. One thus expects this state to be
simply empty AdS space in the classical limit $\ell_p/ \ell_{AdS}
\rightarrow 0$; i.e., $N \rightarrow \infty$.  We shall assume that
this the case in our discussions below.

\section{Unitiarity and AdS/CFT}
\label{unitarity}

Having reviewed some essential features of AdS/CFT,  let us now ask
what this correspondence has to say about unitarity in the quantum
gravity theory.  We will take as given that the CFT is a
well-defined local quantum field theory with a self-adjoint
Hamiltonian. Furthermore, $H$ is one of the operators in ${\cal
D}_{CFT}$ for which we understand the dictionary.  Thus, at least
the boundary observables ${\cal D}_{AdS}$ must evolve unitarity
under the action of the AdS Hamiltonian; i.e., the operators defined
by the asymptotic behavior of bulk fields evolve unitarily.

On the one hand, this statement may seem quite surprising. Unitarity
is of course associated with conservation of information. Suppose
that one creates a state by acting on the vacuum with one of our
boundary observables, perhaps $T_{ij}^{AdS}(y)$.  This essentially
amounts to creating a graviton in the bulk, near the AdS boundary.
This graviton then propagates deeper into the bulk and might  fall
into a black hole. How then can the information in this graviton
still be available on the boundary at later times?  Here it is
especially interesting to consider the unitary evolution over {\em
short} periods of time, too short, say, for there to be any
possibility for the information to return to the boundary in Hawking
radiation.  The unitarity of the CFT seems to suggest that this
information must nevertheless remain present in operators in ${\cal
D}_{AdS}$ at {\em any} later time.

On closer examination, however, this statement need not be a
surprise after all.  In fact, it has been recently argued that such
properties naturally arise in any theory of quantum gravity
\cite{unitarity,ThoughtExp}.  The essential point is that, as
already noted above, the gravitational Hamiltonian is a pure
boundary term, and thus lies in the algebra of boundary observables.
This Hamiltonian generates time translations on the AdS boundary
that are precisely the image under $\phi_{AdS/CFT}$ of CFT
time-translations. Furthermore, while we have little control over
the Hamiltonian of non-perturbative quantum gravity, one may check
\cite{unitarity} that this Hamiltonian is a self-adjoint operator at
each order in perturbation theory, {\em even about a black hole
background}. I.e., there appear to be no obstacles to this
Hamiltonian remaining self-adjoint in the full quantum theory

So then, it appears that information encoded in our graviton is
simultaneously available at two different locations: at the point
deep in the bulk (perhaps within a black hole) to which the graviton
has traveled, and also at the boundary.  We refer the reader to
\cite{ThoughtExp} for a detailed discussion of just how the desired
information might be recovered by a suitable boundary observer, and
for a resolution of what might seem to be potential paradoxes. As
explained in \cite{unitarity}, there is no claim that this
information has in any way been copied into duplicate qubits.
 Such a duplication would violate the quantum `no xerox
theorems' \cite{noxerox}. Instead, there remains a single qubit of
quantum information, but one finds that both boundary operators and
operators deep in the bulk can be sensitive to the same qubit. The
reader should consult \cite{unitarity,ThoughtExp} for further
details.

As a final comment, it is interesting to note that no issues of
`baby universes' (see e.g. \cite{Ted}) arose in our discussion
above.  Since the argument for unitarity depends only on the region
near the boundary, it is independent of whether or not baby
universes form in gravitational collapse far from the boundary.
Suppose in particular that baby universe do exist, that they can be
present in the initial quantum state, and that they contain
additional degrees of freedom not present in the original asymptotic
region. The conclusion of the above argument is that these new baby
universe degrees of freedom simply do not mix with boundary
observables under the boundary time evolution.  This is reminiscent
of the superselection effects noted in certain other discussions of
baby universes \cite{3Qsupsel}.

So, in retrospect, much of what AdS/CFT has to say about unitarity
follows directly from natural extrapolations of bulk gravitational
physics. Does AdS/CFT teach us anything new?  It most certainly
does, in at least three ways. First, it confirms the above
extrapolations.  Second, it tells us about the microscopic density
of states which, when combined with unitarity, makes additional
interesting predictions (see \cite{JM,FL,IKLL,IOP}).  We will
briefly discuss this density of states in section \ref{entropy}.
Third, in specific examples AdS/CFT tells us how to construct
additional complete sets of boundary observables. We argued in
section \ref{BO} that the algebra ${\cal D}_{CFT}$ is likely to be
complete simply because it contains the stress tensor (and perhaps a
few other fields charged under certain symmetries). However, when
the CFT is a gauge theory, another complete set of observables
immediately suggests itself: spacelike Wilson loops at each time. As
noted above, these operators can also be translated to the AdS side
of the correspondence using stringy degrees of freedom (see
\cite{WL}). One thus learns that such stringy boundary observables
at each time are sufficient to generate ${\cal D}_{AdS}$.

\section{The Bekenstein-Hawking Entropy in AdS/CFT}

\label{entropy}

Having reviewed the basics of the AdS/CFT correspondence, and after
our brief aside on unitarity, we may now return to the main question
raised in the introduction:  What, precisely, does AdS/CFT have to
say about the entropy of black holes?

We first note that our unitarity discussion above suggests how the
Hawking radiation cycle example from the introduction might be
reconciled with the idea that black holes have a finite number of
internal states given by $S_{BH}$. Consider the version in which we
start with empty AdS space and proceed by applying various boundary
operators.  We have seen that any information sent into the AdS
space through the boundary does, at least in a certain sense, remain
accessible on the boundary. Thus, there is simply no sharp division
between information ``inside the black hole" and information
``outside."  In particular, boundary operators {\em can} act on
qubits associated with information sent into the black hole.  By
their very nature as annihilation operators, their action can
decrease the space of possible black hole states.  Thus, instead of
increasing each time, the number of states associated with a black
hole of given mass remains the same in each cycle of the experiment.

We can now address what AdS/CFT tells us about entropy.  The basic
ingredients of this story are simply the ability to count states in
the CFT, and the map $\phi_{AdS/CFT}$ that takes states and
observables from one theory to the other.  Performing a precise
counting of CFT states is generally difficult in strongly coupled
CFTs, though simple estimates give a number of states in rough
agreement with the Bekenstein-Hawking entropy of AdS black holes.
The good news is that one {\em can} do a precise counting (in the
limit of large excitations) for 1+1 CFTs which satisfy a property
known as modular invariance.  Since the 1+1 CFTs that arise in
AdS${}_3$ dualities turn out to satisfy this property, Cardy's
formula \cite{Cardy} gives the leading growth of the entropy with
energy and angular momentum:
 \be
 \label{Cardy}
S(E,J) = 2 \pi \sqrt{c(\ell_{AdS} E+J)/12} + 2 \pi \sqrt{\bar
c(\ell_{AdS} E - J)/12}.
 \ee
Here $c$ and $\bar c$ are respectively the left- and right-moving
central charges. As described in \cite{Strominger}, \eqref{Cardy}
agrees precisely with the Bekenstein-Hawking entropy of the
corresponding BTZ black holes in AdS${}_3$.  We will therefore
assume that, if a similarly precise counting of CFT states were
possible in other dimensions, the results would again agree with the
Bekenstein-Hawking entropy of the relevant black hole so long as i)
the semi-classical gravity approximation is valid and ii) one is at
appropriate $E,J$, or other charges so that $S_{BH}$ is much larger
than the entropy of other black holes or of non-black hole states.

So then, what does this counting tell us about quantum gravity
states on the AdS side of the correspondence?   The image of
$\phi_{AdS/CFT}$ in the space of AdS quantum gravity states is
precisely the set of states obtained by acting on the vacuum with
${\cal D}_{AdS}$.  Thus, AdS/CFT tells us that the density of such
black holes states is given by $S_{BH}$.

Now, this statement may not yet appear to be very useful from the
bulk point of view.  After all, we defined ${\cal D}_{AdS}$ to be
the set of boundary operators for which an AdS/CFT dictionary is
understood. What we need to make the statement useful is a more
gravitational characterization of this algebra.   Recall, however,
that we suggested that a sort of ergodicity result might hold in the
CFT that would naturally make ${\cal D}_{CFT}$ equivalent to the
algebra generated by the stress tensor (and perhaps a few other
local observables), and we stated that we would assume this property
below. Since we assume this property to hold for ${\cal D}_{CFT}$,
the map $\phi_{AdS/CFT}$ allows us to carry it over directly to the
AdS side so that it also holds for ${\cal D}_{AdS}$; i.e., it
follows that ${\cal D}_{AdS}$ is the algebra generated by the AdS
boundary stress tensor (and in some cases a few additional boundary
observables).

We now have a precise technical statement, but it may yet convey
little intuition.  Let us therefore recall that, as noted in section
\ref{unitarity}, any AdS operator which might be called a boundary
observable should evolve unitarily under boundary time-translations.
In parallel with our ergodicity assumption for the CFT, it is
natural to expect all such boundary observables to mix with each
other under this evolution\footnote{Again, there are possible
exceptions due to conservation laws.  We shall not mention such
possible exceptions further.  As in the CFT discussion, one expects
such exceptions to be easily dealt with by using a few additional
charged boundary observables to generate the full algebra along with
the stress tensor.}.  This argues that, at least so long as we
consider spacetimes having only a single boundary\footnote{If more
than one boundary is present, the gravity theory will have separate
Hamiltonians generating time translations along each boundary.  See
e.g. \cite{ThoughtExp} for a review of this point.}, it is useful to
think of ${\cal D}_{AdS}$ as the algebra of {\it all} boundary
observables.  In this sense, AdS/CFT states that {\em the
Bekenstein-Hawking entropy gives the density of states created by
acting with boundary observables on the vacuum.}  Or, more
succinctly, one might say that it gives the density of boundary
observable states in the superselection sector defined by the
vacuum.

\section{Discussion: Where is the bag of gold?}
\label{disc}

We have attempted to state precisely what AdS/CFT implies about
black hole entropy.  We found a precise characterization of the
Bekenstein-Hawking entropy as the density of black hole states which
lie in the same superselection sector as the vacuum, where the
notion of superselection sector was defined by a certain observable
algebra ${\cal D}_{AdS}$. This ${\cal D}_{AdS}$ can in turn be
defined as the algebra generated by the boundary stress tensor (and
perhaps a few additional operators), and we argued that it is best
thought of as the algebra of ``all boundary observables."

Where then, does this leave Wheeler's bag of gold and its kin?
Semiclassical considerations suggest that the density of such states
is far higher than the Bekenstein-Hawking entropy.  There are thus
two logical alternatives: i) the semi-classical states do not all
correspond to distinct, well-defined states of the full quantum
gravity theory or ii) these states (or at least, most of them) are
not in the boundary-observable superselection sector containing the
vacuum.   A survey of AdS/CFT researchers would no doubt find
supporters both of (i) and of (ii), though option (ii) seems to have
been more explicitly discussed in the AdS/CFT literature; see e.g.
\cite{JM} in the limit where the internal FRW space becomes and
asymptotically flat spacetime and \cite{FHMMRS} for the case where
the FRW spacetime is de Sitter space.

We also find option (ii) to be more plausible and focus on this
alternative below. In particular, as noted above, at the classical
level no mechanism is known to create such a bag of gold from empty
AdS space by acting with boundary observables. However,  because
bags of gold can classically be smoothly deformed to configurations
which {\em can} clearly be created by the action of boundary
operators, one might naturally ask why bag-of-gold states cannot
arise from the action of boundary operators via some quantum
tunneling process.   Indeed, instantons which appear to describe
tunneling to bag-of-gold-like states exist \cite{FGG,FMP}.  However,
since they are not smooth the issue remains unclear.   I have no
clear answer to this question, other than that AdS/CFT appears to
predict that such tunneling is not possible and that understanding
this prediction from the AdS gravity point of view remains an
important open problem.

In some sense, option (ii) implies that the full quantum gravity
theory contains additional observables not found in ${\cal D}_{AdS}$.
I.e., the algebra of boundary observables is {\em not} complete.
However, it is worth contemplating this statement in more detail.
Consider, for example, the special case in which the FRW region in
the bag of gold is replaced by a second asymptotically anti-de
Sitter region. It is no surprise that such regions cannot be created
by boundary observables associated with the original boundary.  In
fact, it would not surprise a semi-classical physicist to learn that
such states are not part of the same Hilbert space as spacetimes
with a single asymptotic region.  Because they correspond to
different boundary conditions, they are naturally associated with
different classical phase spaces and thus with different quantum
Hilbert spaces.  One might say that they correspond to different
superselection sectors of the theory.  We have also seen that, with
respect to the algebra of boundary observables, the bag of gold
spacetimes also lie in a new superselection sector.  So then, might
it be reasonable to treat finite-sized bags of gold as being on a
similar status to a second asymptotically AdS region?  That is,
might one be able to {\em define} a reasonable theory of quantum
gravity with what amounts to some quantum generalization of a
boundary condition that excludes bags of gold?   The idea is that,
in analogy with boundary conditions, other definitions would exist
for which bags of gold would in fact be allowed, or perhaps even
required, but that nevertheless bag-free definitions would remain
possible.

A detailed analysis of this question would take us far beyond the
scope of this work.  However, the existence of such a definition
seems quite reasonable on the basis of known semi-classical physics.
We have already noted that we expect no process of throwing in particles
through the boundary to create such a bag of gold.  In particular,
even if we throw in a very advanced scientist and a well-equipped
laboratory, no action of the scientist described by classical
gravity will create the bag of gold.  Furthermore, as discussed
above, it seems that quantum tunneling also does not create such
states. Thus, it seems plausible that bag-of-gold operators are not
needed for a reasonable self-contained theory of quantum gravity. In
fact, by focussing on quanta which can be thrown in through the
boundary, we have argued that we need only consider states created
from the vacuum by the action of boundary observables.  In other
words, there does appear to be a reasonable, self-contained theory
of asymptotically AdS quantum gravity in which the boundary
observables form a complete set of operators.  In such a theory, the
Bekenstein-Hawking entropy {\em would} count the total number of
black hole states, and bags of gold simply do not arise.

On the other hand, we have not yet addressed those definitions of
the AdS quantum gravity theory that {\em do} allow bags of gold or
their relatives. The current literature \cite{JM,FHMMRS} suggests
that such theories are dual to a product theory, with one factor
being the usual CFT and the other being some new set of degrees of
freedom. The implication is then that the full theory decomposes
into superselection sectors with respect to our boundary
observables, {\em all} of which are isomorphic to the one containing
the vacuum.  In this context one might say that the
Bekenstein-Hawking entropy counts the density of black hole states
in any superselection sector.  Thus, in what some readers may
consider an ironic twist, it may be fair to say that in AdS/CFT the
Bekenstein-Hawking entropy counts ``not the full set of states
describing the black hole interior, but only those states which are
distinguishable from the outside,'' a point of view which has long
been championed within the relativity community.  However, a
detailed discussion of the product theory point of view raises many
questions about the experience of observers who fall into the black
holes, as well as other questions of local physics not readily
expressed in terms of boundary observables.  As is so often the
case, we end with the conclusion that understanding how to describe
such local bulk observations remains one of the most interesting
open questions in studies of the AdS/CFT correspondence, and in
quantum gravity more generally.

\begin{acknowledgments}
This work was supported in part by the US National Science
Foundation under Grant No.~PHY05-55669, and by funds from the
University of California.
\end{acknowledgments}

\end{document}